\documentclass[reprint,
superscriptaddress,
showpacs,
 amsmath,amssymb,
 aps,
 pra,
]{revtex4-2}

\usepackage[english]{babel}
\usepackage{graphicx}
\usepackage{dcolumn}
\usepackage{bm}
\usepackage[pdfstartview=FitH,colorlinks=true,citecolor=blue,linkcolor=blue,urlcolor=blue]{hyperref}
\usepackage[mathlines]{lineno}
\usepackage{xcolor}
\usepackage{tikz}
\usetikzlibrary{patterns}
\usepackage{natbib}
\usepackage{braket}
\usepackage{cleveref}
\usepackage{longtable}
\usepackage{float}
\usepackage{bm}
\usepackage{tikz}
\usetikzlibrary{calc,patterns,decorations.pathmorphing,decorations.markings}
\PassOptionsToPackage{hyphens}{url}

\newcommand{\up}{\uparrow}
\newcommand{\down}{\downarrow}

\newcommand{\Ntwo}{\frac{N}{2}}
\newcommand{\hHs}{\hat{H}_s}
\newcommand{\hP}{\hat{P}}

\usepackage{soul}

\begin{document}

\title{Dynamical probing of high-order spin coherence in one-dimensional mixtures}

\author{S. Musolino}
\affiliation{Universit\'e Grenoble Alpes, CNRS, LPMMC, 38000 Grenoble, France}
\affiliation{Universit\'e C\^ote d'Azur, CNRS, Institut de Physique de Nice, 06200 Nice, France}
\author{G. Aupetit-Diallo}
\affiliation{Universit\'e C\^ote d'Azur, CNRS, Institut de Physique de Nice, 06200 Nice, France}
\affiliation{SISSA, Via Bonomea 265, I-34136 Trieste, Italy}
\author{M. Albert} 
\affiliation{Universit\'e C\^ote d'Azur, CNRS, Institut de Physique de Nice, 06200 Nice, France}
\affiliation{Institut Universitaire de France}
\author{P. Vignolo}
\affiliation{Universit\'e C\^ote d'Azur, CNRS, Institut de Physique de Nice, 06200 Nice, France}
\affiliation{Institut Universitaire de France}
\author{A. Minguzzi}
\affiliation{Universit\'e Grenoble Alpes, CNRS, LPMMC, 38000 Grenoble, France}

\begin{abstract}
We investigate the dynamics of one-dimensional SU(2) ultracold fermions near the Tonks-Girardeau limit, confined in a box potential. The system is driven out of equilibrium by initially preparing the two spin components in a fully separated configuration, and its evolution is described by the Hamiltonian in the presence of strong repulsive interactions. Building on the results in G. Aupetit-Diallo et al., \href{https://journals.aps.org/pra/abstract/10.1103/PhysRevA.107.L061301}{Phys. Rev. A {\bf 107}, L061301 (2023)}, we extend the analysis to out-of-equilibrium dynamics, uncovering the emergence of time-dependent oscillating high-momentum tails in the momentum distribution. These oscillations, due to the finite size of the system, are governed by a nonlocal, high-order spin coherence term, whose amplitude and phase evolve over time. We show that this term initially grows as $t^{N/2}$ and subsequently follows the spin-mixing dynamics of the system. Notably, when the spin components are fully mixed, the amplitude of this border-to-border spin coherence reaches its maximum value.
\end{abstract}

\maketitle

\section{Introduction}\label{sec:intro} 

Strongly interacting one-dimensional (1D) multicomponent systems provide us a handy platform for exploring quantum magnetism without the need for an underlying lattice structure~\cite{art:deuretzbacher2014, art:massignan2015, art:volosniev_2015, art:Yang_PRA2015}. When particles are confined to one dimension under highly repulsive interactions, they enter the so-called fermionized or Tonks-Girardeau (TG) regime~\cite{art:tonks1936,art:Girardeau1960,art:minguzzi2022strongly}. In this regime, the many-body wave function vanishes when two particles approach each other and changes sign according to the exchange symmetry of the particles.  For multicomponent homogeneous~\cite{Oelkers_2006} and inhomogeneous~\cite{art:volosniev_nat}  systems in the presence of strong interactions, the spin and spatial degrees of freedom decouple and  the wave function factorizes in two parts in each coordinate sector associated to the ordering of particles in a one-dimensional geometry:  the spatial part,  behaving as the one of noninteracting  fermions, and the spin part,  governed by an effective spin-chain Hamiltonian~\cite{art:deuretzbacher2008_lett, art:volosniev_nat}. The energy spectrum of these SU($\kappa$) systems, with $\kappa$ being the number of spin components, exhibits a high degree of degeneracy, which is determined by the numerous possible spin configurations~\cite{art:girardeau2007soluble}. The coherent superposition
between the spin configurations can produce states with well-defined or mixed  symmetry under particle exchange~\cite{art:decamp2016exact}, which can be experimentally probed by looking at the zero-momentum peak \cite{art:musolino2024} and the large momentum tails \cite{art:decamp2016_high,art:aupetit} of the momentum distribution.

The zero-momentum peak of the momentum distribution is indicative of quasi long-range order and depends on the sum of all-order spin correlations, weighted by the inverse square root of spin distance~\cite{art:forrester2003,art:aupetit}. At asymptotically large momenta, the behavior of the momentum distribution depends only, except in particular cases~\cite{art:corson_bohn2016, art:bouchoule_Tan2021, art:schehr_wigner2021, art:Rylands2023, art:aupetit2023, art:derosi2024}, on the many-body wave function in the limit of vanishing relative distance between two atoms and, therefore, only on nearest neighbor spin correlations. In this limit, for contact interactions, the derivative of the wave function is discontinuous and this leads to a characteristic power-law decay in the momentum distribution, $n(k) \sim 1/k^4$ \cite{art:minguzzi_PLA2002,art:olshanii_PRL2003}. The weight of this power-law tail grows as the symmetry of the many-body wave function increases~\cite{art:decamp2016_high} and is  related 
 by one of the Tan relations~\cite{art:tan2008energetics, art:Tan_largemom, art:tan2008generalized} to the  so-called Tan contact, which is proportional to the derivative of the total energy of the gas with respect to the inverse of the interaction strength~\cite{art:Tan_largemom}. 

In particular, as recently studied by some of us~\cite{art:aupetit2023}, the presence of hard-wall confinements introduces additional terms in the asymptotically large momentum behavior of the momentum distribution. Those terms, an offset and an oscillating term, derive from the singularity of the wave function at the edges of the trap, which   also gives rise to a  $1/k^4$ decay. Interestingly, in atomic mixtures, the amplitude of these oscillations is nontrivial and depends on a highly nonlocal spin coherence term.
Specifically, this coherence involves spin states connected by a cyclic permutation of the spin at one edge of the trap through all other spins  to the opposite edge. As a result, the border-to-border coherence is highly sensitive to the specific state of the mixture~\cite{art:aupetit}, and, as presented in this work, its time evolution is intricately linked to the underlying spin ordering.

In this article, we extend the analysis of Ref.~\cite{art:aupetit2023} to a time-dependent scenario by focusing on SU$(2)$ fermions confined in a box potential. We show that the tails of the time-dependent momentum distribution give direct access to the nonlocal high-order spin correlations, $c_\sigma^{(1, N)}(t)$,  where $(1, N)$ indicates the $N$-cycle connecting  the first and the $N$-th particle in the box.
We follow its time evolution starting from an initially spin-separated state, as recently investigated in Refs.~\cite{art:pecci2022, art:musolino2024, art:patu2024numerical}.
We demonstrate that $c_\sigma^{(1, N)}(t)$ initially grows as $t^{N/2}$ with $N/2$ corresponding to the minimum number of exchanges necessary to bring the spin at the interface to the furthest border, and this result depends on the degree of blending of  the initial configuration.  Furthermore, we show that at later times the nonlocal spin correlation 
follows the dynamics of the induced magnetization and becomes larger when the average magnetization is approximately zero. Our results are valid also for bosons with different values for intra- and interspecies interactions, where the breaking of the SU($2$) symmetry introduces an additional oscillation in time~\cite{art:musolino2024}, and can be  generalized  to mixtures with $\kappa > 2$ components.

The article is organized as follows. In Sec.~\ref{sec:model}, we present our model Hamiltonian, the definition of the correlation functions, and the relation with the spin-chain model. In Sec.~\ref{sec:res}, we present our numerical results for few-particle systems from $N=4$ to $10$ particles by dividing the analysis   of the nonequilibrium dynamics in the very early times (Sec.~\ref{sec:HOSC}) and later times (Sec.~\ref{sec:tlonger}). In Sec.~\ref{sec:end}, we summarize our results and discuss the possible applications for experiments and other theoretical perspectives.   We finally present the details of the derivations in Appendices~\ref{app:k4nk}, ~\ref{app:corr_tshort}, and~\ref{app:d0}.

\section{Model}
\label{sec:model}
The Hamiltonian for a 1D two-component ($\sigma =\uparrow$ or $\downarrow$) mixture of $N$ fermions with mass $m$ trapped in a hard wall box potential of size $L$ and interacting via a two-body contact interaction of strength $g$ is given by
 \begin{equation}
 \begin{split}
H&=\!\sum_{i=1}^{N}
 \Big[\!\!-\frac{\hbar^2}{2m}\frac{\partial^2}{\partial x_{i}^2} + V(x_i)\Big]\!+g \sum_{i=1}^{N_{\uparrow}} \sum_{j=N_\uparrow+1}^{N}\delta(x_{i}-x_{j}),
 \end{split}
 \label{eq:ham}
 \end{equation}
 where $V(x) = 0$ for $x \in [-L/2, L/2]$ and $\infty$ otherwise.  In the limit $g\rightarrow +\infty$, the time-dependent many-body wave function of this system, $\Psi(\vec{X}, \vec{\sigma}, t)$ with positions $\vec{X} = \{x_1, x_2, \dots, x_N \}$ and spins $\vec{\sigma} = \{\sigma_1, \sigma_2, \dots, \sigma_N \}$, vanishes whenever $x_i=x_j$. In the presence of spin excitations only, the latter can be written as follows~\cite{art:girardeautimedependent, art:deuretzbacher2008_lett, art:volosniev_nat}:
 \begin{equation}
  \Psi(\vec{X}, \vec{\sigma}, t)= \sum_{P\in S_N}   a_P(t) \theta_P(\vec{X})\Psi_A(\vec{X}),
  \label{eq:Psi_MB}
  \end{equation}
  where the $\vec{\sigma}$-dependence is encapsulated in the spin coefficients $a_P(t)$ with $P$ representing a permutation of $N$ elements in the symmetric group $S_N$. The function $\theta_P(\vec{X})$ is the generalized Heaviside function, equal to $1$ in the coordinate sector $x_{P(1)}<\dots<x_{P(N)}$ and $0$ elsewhere, and $\Psi_A=A \Psi_F$ 
  with $A =\prod_{i<j} \mathrm{sgn}(x_i-x_j)$, and $\Psi_F=(1/\sqrt{N!}) \det [\phi_j(x_k)]$ is the Slater determinant built from the natural one-particle orbitals of the box, $\phi_j(x)$. For convenience, we use a symmetrized wave function as building block for the mixture solution, so that the sign taking into account for antisymmetric exchanges is indicated explicitly in the following expressions. 

At strong coupling, in the limit $1/g \rightarrow 0$,
the coefficients $a_P(t)$ of Eq.~\eqref{eq:Psi_MB} for the ground-state manifold
are determined  by considering the expansion of the total energy of the system $E \simeq E_A - K/g$, 
where $E_A$ is the Fermi energy associated to $\Psi_A$ and $K=K(a_0, a_1, \cdots)$ is a functional of the coefficients $a_P$~\cite{art:volosniev_nat}. The energy eigenvectors and eigenvalues of the system can be  derived by varying $K$ with respect to the coefficients $a_P$ ($\partial K /\partial a_P = 0$) and diagonalizing the resulting matrix.

In the limit of $1/g \to 0$, the Hamiltonian in Eq.~\eqref{eq:ham} can be mapped  onto the following spin-chain Hamiltonian~\cite{art:deuretzbacher2014, art:volosniev_2015, art:massignan2015, art:Yang_2016}:
\begin{equation}
\begin{split}
\hat{H}_s &=\left(E_A - \sum_{i=1}^{N-1} \frac{\alpha_i}{g}\right)\mathbf{1}  + \sum_{i=1}^{N-1} \frac{\alpha_i}{g} \hat{P}_{i, i+1} \\
&= E_A \mathbf{1}  + 2\sum_{i=1}^{N-1} \frac{\alpha_i}{g} \left(\mathbf{S}_{i}\cdot\mathbf{S}_{i+1} - \frac{1}{4}\right),
\end{split}
\label{eq:Hspin}
\end{equation}
where $\mathbf{1}$ is the identity matrix, $\hat{P}_{i, i+1}$ is the nearest-neighboring permutation operator, $\mathbf{S}_i =(S_i^{(x)}, S_i^{(y)}, S_i^{(z)})$ are the spin operators of the $i$-th particle,
and the exchange coefficients are given by
\begin{equation}
\begin{split}
\alpha_i &= \frac{N! \hbar^4}{m^2 } \int d\vec{X} \theta_\mathrm{id}(\vec{X})\delta(x_i - x_{i+1})\left|\frac{\partial \Psi_A}{\partial x_{i}}\right|^2.
\end{split}
\label{eq:Ji}
\end{equation}
To obtain the second line of  Eq.~(\ref{eq:Hspin}), we have used the relation $\hat{P}_{i, i+1}=2\mathbf{S}_{i}\cdot\mathbf{S}_{i+1} + 1/2$, which is specific to SU(2) systems.
We underline  that Eq.~(\ref{eq:Hspin}) has the same symmetries of the original Hamiltonian (\ref{eq:ham}), namely it is symmetric under time-reversal (spin flip) and parity, for which $\alpha_i=\alpha_{N+1-i}$.

\subsection{One-body density matrix and momentum distribution}
\label{sec:corr}
The one-body density matrix of 
 the $\sigma=\uparrow, \downarrow$ component of the strongly interacting mixture under consideration can be decomposed as~\cite{art:deuretzbacher2016_num}
\begin{equation}
\rho^{(1)}_\sigma(x, x', t) = \sum_{i, j=1}^N (-1)^{i+j} c_\sigma^{(i, j)}(t) \rho^{(i, j)}(x, x'),
\label{eq:rho1}
\end{equation}
where the $(-1)^{i+j}$ factor  is present because the mixture is fermionic. 
The spatial
part, $\rho^{(i, j)}(x, x')$, is given by 
\begin{equation}
\begin{split}
\rho^{(i, j)} (x, x')&= N! \displaystyle\int_{I_{ij}} \prod_{n \neq i} dx_n \Psi^*_\mathrm{A}(\cdots, x_{i-1}, x, x_{i+1}, \cdots)\\
&\times\Psi_\mathrm{A}(\cdots, x_{i-1}, x', x_{i+1}, \cdots),
\end{split}
\label{eq:rhoij}
\end{equation}
where $I_{ij}$ is the integration interval constrained by $x_1 < \cdots < x_{i-1} < x < x_{i+1} < \cdots < x_{j-1} < x' < x_{j} < \cdots < x_N$. The spin part, $c_\sigma^{(i, j)}(t)$, is given by~\cite{art:aupetit2023} 
\begin{equation}
\begin{split}
    c_\sigma^{(i,j)} (t)&=\delta_{\sigma, \sigma_i} \sum_{P\in S_N} a_{P}^\ast(t) a_{P_{(i, \dots, j)}}(t)\\
     &= \sum_{\vec{\sigma}\in \Sigma} \delta_{\sigma, \sigma_i} \braket{\chi(t) | \vec{\sigma}} \braket{\vec{\sigma} |\hat{P}_{(i, \dots, j)} |\chi(t)},
         \end{split} 
    \label{eq:cij_single}
\end{equation}
where $\ket{\chi(t)} = e^{-i\hat{H}_st}\ket{\chi(0)}$ is a generic time-dependent spin state evolving under the action of the Hamiltonian in Eq.~\eqref{eq:Hspin} from the initial state $\ket{\chi(0)}$, $\Sigma$  denotes the spin space of size $(N!/N_\up!N_\down!)$ containing all the possible spin-ordering configurations (snippets), and $\hat{P}_{(i, \dots, j)}$ is a cyclic (anticyclic) permutation from $i$ to $j$ if $i<j$ ($i>j$) and the identity if $i=j$~\cite{art:deuretzbacher2016_num}.  We notice that the Hamiltonian $\hat{H}_s$ is not diagonal in the basis of the states $\ket{\vec{\sigma}}$. The $c_\sigma^{(1, N)}(t)$ correlation evaluates the overlaps
between states that have the same sequence of $N-1$ spin orientations inside the chain, while they differ by exchanging the first spin, if its spin is $\sigma$, with the last one. In other words, these states are brought one onto the other by a $N$-cycle permutation. 

 To clarify the connection with spin operators,  we can rewrite the total $c^{(i, j)}(t) = \sum_\sigma  c^{(i, j)}_\sigma(t) = \bra{\chi(t)}  \hat{P}_{(i, \dots, j)} \ket{\chi(t)}$   in terms of spin operators, which reads as, for $i< j$,
\begin{equation}
\begin{split}
c^{(i, j)}(t) 
 &=\bra{\chi(t)}  \Big(2\mathbf{S}_{i} \cdot \mathbf{S}_{i+1} +\frac{1}{2}  \Big) \prod_{k=i+1}^{j-2} \Big(2\mathbf{S}_{k} \cdot \mathbf{S}_{k+1} \\&+ \frac{1}{2} \Big)   \Big(2\mathbf{S}_{j-1} \cdot \mathbf{S}_{j} +\frac{1}{2}  \Big) \ket{\chi(t)},
\end{split}
\label{eq:cij_spin}
\end{equation}
where we have used that any cyclic permutation can be decomposed into adjacent transpositions,  i.e., $\hat{P}_{(i, \dots, j)}  = \hat{P}_{i, i+1} \hat{P}_{i+1, i+2} \cdots \hat{P}_{j-1, j}$. We notice that Eq.~\eqref{eq:cij_spin} is consistent with the spin contribution to the momentum distribution of the one dimensional strongly interacting Hubbard model presented in Ref.~\cite{art:ogata_shiba}.

Remarkably, the element $(1, N)$ of Eq.~\eqref{eq:cij_spin} does not represent a two-point correlation function between the two most distant spins, but also encodes information about the string of spins of length $N-2$ that separates the spins at positions $1$ and $N$, making it a higher-order spin correlation. This is reminiscent of the so-called string order in spin chains~\cite{art:denNijs_SO, art:Kennedy_SO}.

The momentum distribution of the spin component $\sigma$ is defined as
\begin{equation}
n_\sigma(k, t) = \dfrac{1}{2\pi} \displaystyle\int dx \int dx' \rho^{(1)}_\sigma(x, x', t)e^{-i k (x-x')}.
\label{eq:nk_sigma}
\end{equation}
As analyzed in Ref.~\cite{art:aupetit2023}, the large-$k$ behavior of the momentum distribution
in the presence of a hard-wall confinement includes two distinct contributions:  The first arises from the Tan contact~\cite{art:tan2008energetics,art:tan2008generalized, art:Tan_largemom}, which is related to two-body correlations and contact interactions; the second represents finite-size effects, vanishing in the thermodynamic limit ($N, L \to \infty$ with $N/L$ finite). Interestingly, this second term remains nonzero even for noninteracting particles~\cite{art:aupetit2023, art:schehr_wigner2021}.

These two contributions arise from different limiting behaviors of  Eq.~\eqref{eq:rho1}:  The first corresponds to the limit where the relative distance between two particles vanishes ($|x-x'| \to 0$), while the second  comes from the behavior near the system's boundaries, $x, x' \to \pm L/2$. Indeed, due to contact interactions, in the region where two particles approach each other ($x \to x'$), the mapping function $\Psi_A(\vec{X})$ entering Eq.~(\ref{eq:Psi_MB})
behaves as
 \begin{equation}
\Psi_A(\vec{X}) \stackrel{|x - x'|\to 0}{\approx} |x-x'| \Phi_0 \left(\frac{x+x'}{2}, x_n \neq \{x, x'\}\right),
\label{eq:PsiA_exp}
\end{equation}
where  $|x - x'|$ is the two-body wave function satisfying the two-body Schr\"odinger equation in the TG limit and $\Phi_0$ is a regular function of the center of mass of $x$ and $x'$ and the rest of the coordinates~\cite{art:olshanii_PRL2003}. Equation~\eqref{eq:PsiA_exp} implies that the wave function has a discontinuous derivative (a cusp) at $x = x'$. 

In the case of hard-wall confinement, a similar discontinuity arises near the system boundaries, where we can write
 \begin{equation}
\Psi_A(\vec{X}) \stackrel{x \to \pm L/2}{\approx} \left|x \mp \frac{L}{2}\right| \Phi_L(x_2, \dots, x_N),
\label{eq:PsiA_exp_single}
\end{equation} 
 with $\Phi_L$ a regular function of the remaining $N-1$ coordinates. Since the wave function is not defined for $x>L/2$ ($x < -L/2$), Eq.~\eqref{eq:PsiA_exp_single} gives only half of a cusp.

In the large-$k$ limit of Eq.~\eqref{eq:nk_sigma}, all the discontinuities must be accounted for. To do so, we use the standard procedure based on the Watson lemma~\cite{art:olshanii_PRL2003}, which says that if we consider a generic function  $f(x)$ with a singularity of the type $|x-x_0|^\eta F(x)$  at $x_0$ with $F(x)$ a regular function and $\eta> -1, \eta \neq 0, 2, 4, \cdots$, then $\lim_{k\to \infty}\int dx e^{-ik x} f(x) = 2 \cos(\pi(\eta +1)/2) \Gamma(\eta + 1) \mathrm{exp}(-ik x_0) F(x_0)/|k|^{\eta+1} + o (1/|k|^{\eta +2})$. Using Eqs.~\eqref{eq:PsiA_exp} and~\eqref{eq:PsiA_exp_single} and Watson's lemma, we can write
 \begin{equation}
 \begin{split}
\lim_{k\to \infty}\int dx e^{-ik x} \Psi_A(\vec{X})  = -\frac{2}{k^2} \sum_{x_0}^\mathrm{sing.} e^{-ik x_0} \Phi_{x_0}(\vec{Y}),
\end{split}
\label{eq:watson}
\end{equation} 
where $x_0$ are the singularity points and $\Phi_{x_0}(\vec{Y})$ is the regular function depending on the vector $\vec{Y}$ that contains all the coordinates except the singular one. 
We notice that the many-body wave function $\Psi(\vec{X})$ in Eq.(\ref{eq:Psi_MB}) has less cusps than
$ \Psi_A(\vec{X})$, since, e.g., it changes sign on exchange of fermions belonging to the same spin component.  This property, as well as the final number of cusps, is accounted for by the choice of the weights $a_P$. 

Using  Eq.(\ref{eq:watson}), we obtain for the $\sigma$-component momentum distribution [Eq.~\eqref{eq:nk_sigma}]
\begin{equation}
\begin{split}
\lim_{k \to \infty} k^4 n_\sigma(k, t) &=\mathcal{C}_N^\sigma(t) + \mathcal{B}_N \left(c_{\sigma}^{(1,1)}(t)+c_{\sigma}^{(N,N)}(t)\right) \\&-\mathcal{A}_N |c_{\sigma}^{(1,N)}(t)|2\cos(kL+ \theta_\sigma(t)),
\end{split}
\label{eq:KN_t_sig}
\end{equation}
where the short-distance contribution related to the component $\sigma$, $\mathcal{C}_N^\sigma(t)$, derives  from Eqs.~(\ref{eq:PsiA_exp}) and the border contributions, $\mathcal{B}_N=(N+1)(2N+1)\pi/(6L^3)$, and $\mathcal{A}_N=(-1)^N (N+1)\pi/(2L^3)$  from Eq.~\eqref{eq:PsiA_exp_single}~\footnote{For convenience, we have redefined the coefficients $\mathcal{A}_N$ and $\mathcal{B}_N$ with respect to Ref.~\cite{art:aupetit2023}.}.
The full derivation of Eq.~\eqref{eq:KN_t_sig} can be found in Appendix~\ref{app:k4nk}. In particular, the sum over $\sigma$ of the short-distance terms corresponds to the Tan contact, $ \mathcal{C}_N =\sum_\sigma \mathcal{C}_N^\sigma(t)$, namely
\begin{equation}
    \mathcal{C}_N = - \frac{m^2}{\pi \hbar^4}\braket{ \chi(t) |\frac{\partial \hat{H}_s}{\partial 1/g} | \chi(t)}, 
    \label{eq:Tan_def}
    \end{equation}
 which does not depend on time, due to the SU(2) symmetry of the mixture, and as also verified in Appendix~\ref{subsec:shortdis}. 

 Regarding  the term proportional to $\mathcal{B}_N$, we notice that in case of time-reversed symmetric  initial state, we can use that $c_{\sigma}^{(1,1)}(t) = 1-c_{\sigma}^{(N,N)}(t)$, from which follows that that term is time independent. 
Finally,  the amplitude of the momentum oscillations depends on the long-range spin coherence $c_{\sigma}^{(1,N)}(t)$ [see Eq.~\eqref{eq:cij_single}], which is a complex-valued correlation, whose modulus $|c_{\sigma}^{(1,N)}(t)|$ and phase $\theta_\sigma(t)$ both depend on time.

Another interesting property related to the last term of Eq.~\eqref{eq:KN_t_sig} is
\begin{equation}
c_{\up}^{(1,N)}(t) = [c_{\down}^{(1,N)}(t)]^\ast = |c_{\up}^{(1,N)}(t)| e^{i \theta_\up(t)},
\label{eq:cup_cdown}
\end{equation}
which yields to $|c_{\up}^{(1,N)}(t)|=|c_{\down}^{(1,N)}(t)|=|c_{\sigma}^{(1,N)}(t)|$, and $\theta_\up(t) = - \theta_\down(t)$. This property holds only for initial states, as the ones considered in this work, that verify $\hat{\Pi}\ket{\chi(0)} = \hat{\Theta} \ket{\chi(0)}$, with $\hat{\Pi}$ the parity operator and $\hat{\Theta}$ the spin-flip operator. 
Therefore, for this class of initial states, the oscillations of the two components are in phase when $\theta_\sigma(t)= n \pi$ and in counter-phase when $\theta_\sigma(t)= (2n+1) \pi/2$ with $n=0, 1,2, \dots$, as we will discuss in Sec.~\ref{sec:tlonger}.

\section{Results}
\label{sec:res}

In this section, we present the numerical results of the dynamics of few-particle systems at a fixed density, $N/L$, with $N = 4, 6, 8, 10$ particles. The initial state is prepared in the nonequilibrium configuration $\ket{\chi(0)} = \ket{\up \up \cdots \up \down \cdots \down \down}$, using a finite but large interaction strength of $g = 20(\hbar^2/mL)$~\footnote{The numerical value of $g$ dictates the time scale of the system  and a different choice simply leads to a rescaling factor that does not alter the results of this work.}.
 Once the dynamics begins, the spin part of the wave function, as described in Eq.~\eqref{eq:Psi_MB}, evolves according to $\ket{\chi(t)} = e^{-i\hat{H}_s t} \ket{\chi(0)}$, while the orbital part remains static. This is due to our choice of the dynamical protocol, which does not involve dynamical variations of the trapping potential.

The rest of the section is divided into two parts: We  examine first the initial build up of the $c_\sigma^{(1, N)}(t)$ correlations (Sec.~\ref{sec:HOSC});  we discuss then the dynamics related to the oscillations of the average magnetization of the system (Sec.~\ref{sec:tlonger}). 

\subsection{Initial growth of the high-order spin coherence}
\label{sec:HOSC}
As discussed in Sec.~\ref{sec:corr},  the nonlocal spin coherence $c_\sigma^{(1, N)}(t)$ represents a many-body correlation. Here, we demonstrate that its initial growth depends on the complexity of the initial state, i.e., the number of spin domains. For the initial state $\ket{\chi(0)} = \ket{\up \up \dots \up \down \dots \down \down}$ with a single spin domain, we observe that the modulus of $c_\sigma^{(1, N)}(t)$ initially grows as $t^{N/2}$. This power-law exponent can be understood as  
the minimum number of spin exchanges required for the spin at the interface 
to reach the opposite edge of the trap (see also Appendix~\ref{app:corr_tshort}).

By performing the Taylor expansion around $t=0$ of Eq.~\eqref{eq:cij_single} for $(i, j)=(1, N)$, we obtain
\begin{equation}
    \begin{split}
    c_\up^{(1, N)}(t) &= \sum_{\beta_1, \beta_2 =0}^\infty \frac{1}{\beta_1 ! \beta_2 !} \sum_{\vec{\sigma} \in \Sigma} \delta_{\up, \sigma_1} 
    \braket{\chi(0)|   (i\hHs t)^{\beta_1}|\vec{\sigma}}\\ 
    &\bra{\vec{\sigma}} \hat{P}_{(1, \dots, N)}  (-i\hHs t)^{\beta_2}\ket{\chi(0)},
    \end{split}
    \label{eq:c1N_sigma_Taylor}
\end{equation} 
where the spin Hamiltonian involves only transpositions between neighboring spins [see Eq.~(\ref{eq:Hspin})]. If we consider $\ket{\chi(0)} = \ket{\up \up \dots \up \down \dots \down \down}$, then  we obtain that the  lowest-order contribution in Eq.~\eqref{eq:c1N_sigma_Taylor}  that is nonzero is 
\begin{equation}
    \begin{split}
    c_{\up}^{(1, N)}(t)
    &\approx \frac{(-i t)^{N/2}}{(N/2)!} \bra{\up \up \dots \up \down \dots \down \down} \hat{P}_{(1, \dots, N)} \\
     &\hHs^{N/2} \ket{\up \up \dots \up \down \dots \down \down} ,
    \end{split}
    \label{eq:c1N_sigma_tapprox0}
\end{equation}
where $\beta_1=0$ and $\beta_2=N/2$  (see Appendix~\ref{app:corr_tshort} for details).  

\begin{figure}
    \centering
   \includegraphics[scale=0.7]{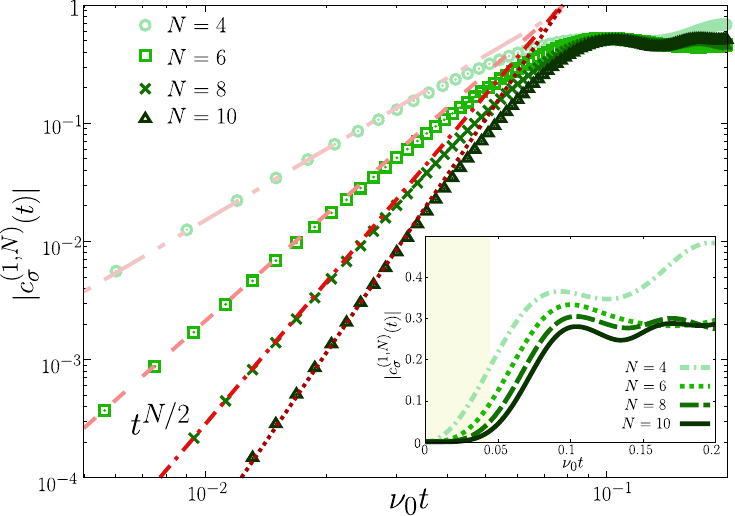}
    \caption{Initial growth of the amplitude of the single-component spin correlation $c^{(1,N)}_\sigma(t)$ for $\sigma = \up$ and $\down$ [see Eq.~\eqref{eq:cij_single}] for $N=4, 6, 8, 10$ in log-log scale. The dashed lines represent the fitting functions that behave as $\sim t^{N/2}$. Inset: Same as the main figure without log-log scale. The yellow area indicates the region where the power-law $t^{N/2}$ holds. The time is in units of $1/\nu_0$ with $\nu_0 = \hbar/(m L^2)$. }
    \label{fig:c1N_sigma}
\end{figure}

To understand the result in Eq.~\eqref{eq:c1N_sigma_tapprox0}, there are two important factors to take into account in Eq.~\eqref{eq:c1N_sigma_Taylor}: The  $\delta_{\up, \sigma_1}$, which dictates that the spin at position $1$ of the configuration must be $\up$, and the action of $\hat{P}_{(1, \dots, N)}$ on the ket state $\ket{\up \up \cdots \up \down \cdots \down \down}$, which consists of moving the $N$th spin to position $1$. Therefore, to have a nonzero result in the right-hand Taylor expansion, $\hHs$ must be applied a $\beta_2$ number of times such that  a spin $\up$ is brought to position $N$. Since $\hat{H}_s$ acts through nearest-neighboring exchanges and for the  selected initial state the spin $\up$ closest to the right edge is at the interface (at position $N/2$),  it is necessary to apply the operator $\hat{H}_s$ at least $N/2$ times. By doing that, we obtain
\begin{equation} 
\begin{split}
&\bra{\vec{\sigma}}\hat{P}_{(1, \dots, N)} (\hat{H}_s)^{N/2} \ket{\up \up \dots \up \down \dots \down \down} \\
&= \braket{\vec{\sigma} | \up \up \dots \up \down \dots \down \down}, 
\end{split}
\label{eq:P1N_Hs} 
\end{equation} 
namely, $\beta_2=N/2$ and on the left-hand Taylor expansion of Eq.~\eqref{eq:c1N_sigma_Taylor} the lowest order that contributes has $\beta_1=0$. Figure~\ref{fig:c1N_sigma} presents the numerical results for the evolution of the modulus of $c_\sigma^{(1, N)}(t)$ with increasing particle numbers $N = 4, 6, 8, 10$ and, as expected, the initial growth follows a $t^{N/2}$ behavior.

As evident from the proof above and discussed with more details in Appendix~\ref{app:corr_tshort}, this result is highly dependent on the structure of the initial state. For an  initial state with more spin domains, such as a state with N\'eel order, $\ket{\chi(0)}= \ket{\up \down \up \down \dots \up \down}$, we find instead that $c^{(1, N)}_\sigma(t) \sim t^{N-1}$ at $t\approx 0$. This suggests that the early-time dynamics of the spin coherence term serve as an indicator of the structural complexity of the initial state.

\subsection{Correlation oscillations and spin-mixing dynamics}
\label{sec:tlonger}
\begin{figure}
    \centering
   \includegraphics[scale=0.85]{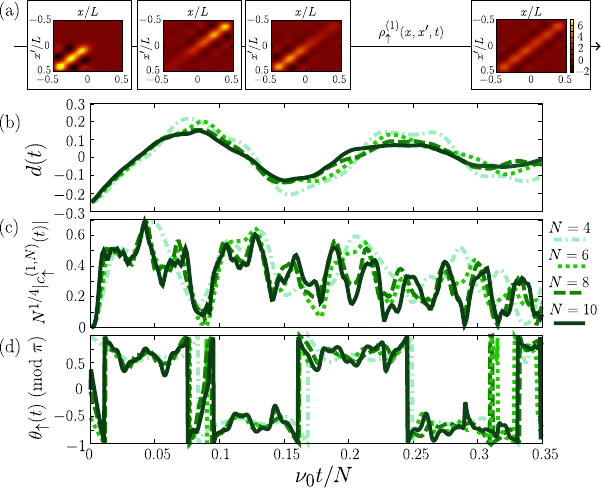}
    \caption{Time evolution from the initial state $\ket{\up \up \dots \up \down \down \dots \down}$ for systems with fixed density $N/L$ with $N$ the number of particles and $L$ the size of the box potential. (a) Snapshots of the $\uparrow$-component one-body density matrix $\rho^{(1)}_\up (x, x', t)$  at fixed times: $\nu_0 t/N = 0, 0.08, 0.15, 0.35$ (from left to right, respectively) for $N=6$  with $x, x'$ in units of $L$. (b) Center of mass displacement of the magnetization $d(t)$ [Eq.~\eqref{eq:dt}]; (c) rescaled amplitude of the single-component spin correlation, $|c_\up^{(1, N)}(t)|= |c_\down^{(1, N)}(t)|$; and (d) phase of $c_\up^{(1, N)}(t)$, $\theta_\up(t) = - \theta_\down(t)$ [see Eq.~\eqref{eq:KN_t_sig}] as a function of the rescaled time $\nu_0 t/N$. }
    \label{fig:dt_sig}
\end{figure}

We now turn to the dynamics at longer timescales. To track the real-space evolution of systems with strong interactions, starting from the initial state $\ket{\up \up \dots \up \down \down \dots \down}$, we monitor the center-of-mass oscillations of the magnetization~\cite{art:pecci2022}
\begin{equation}
d(t) = \displaystyle\frac{1}{N}\displaystyle\int_{-L/2}^{L/2} dx\,x\, m(x, t),
\label{eq:dt}
\end{equation} 
where $m(x, t) = n_{\uparrow}(x, t) - n_{\downarrow}(x, t)$ is the local magnetization and $n_{\sigma}(x, t)= \sum_{i=1}^N (\sum_{P} |a_P(t)|^2 \delta_{\sigma, \sigma_i})\rho^{(i, i)}(x, x)$ is the density of a single component. The zeros of $d(t)$ indicate the times when the center of mass of the magnetization is at the center of the trap, and the oscillations of $d(t)$ reveal the dynamical imbalance between the single-component densities, as depicted in Figs.~\ref{fig:dt_sig}(a)-(b).  

At $t=0$, the $\up$ component of the one-body density matrix $\rho^{(1)}_\up (x, x', 0)$ is nonzero only in one side of the box [$-L/2 < x, x'< 0$ in Fig.~\ref{fig:dt_sig}(a)]. 
During the initial evolution, each component $\rho^{(1)}_\up (x, x', t)$ starts oscillating from one side to the other until after a certain time ($\nu_0 t/N \gtrsim  0.35 $) it is well distributed in all the space. For the box with length $L$ scaling with $N$,  we found that the period of $d(t)$ is proportional to $1/N$, as shown in Fig.~\ref{fig:dt_sig}(b). Notice that this scaling is not visible in Fig.~\ref{fig:c1N_sigma} since it builds up  at longer timescales.
The scaling found here differs from the one found in Ref.~\cite{art:pecci2022}, since in the latter case the system was trapped in a harmonic potential and the number of particles was changed while keeping fixed the harmonic trap frequency.

Figures~\ref{fig:dt_sig}(c)-(d) present the evolution of  $c_\up^{(1, N)}(t)$ in terms of its amplitude (c) and its phase (d).  We remind that, according to Eq.~\eqref{eq:cup_cdown}, the amplitude of $\down$ component is equal to the one of the $\up$ component, $|c_\down^{(1, N)}(t)| = |c_\up^{(1, N)}(t)|$, and its phase is equal in modulus and opposite in sign, $\theta_\down(t) = - \theta_\up(t)$. At $t=0$, there is no correlation between opposite sides of the trap, leading to  $c_\up^{(1, N)}(0)=0$. Once the dynamics begins, particles oscillate within the trap, and  $c_\up^{(1, N)}(t)$ builds up while the two spin components mix. The correlation 
reaches its maximum when both components are distributed evenly across the entire trap, corresponding to the time  where $d(t) \approx 0$.

As discussed in Sec.~\ref{sec:HOSC}, during the early dynamics (where $|c_\up^{(1, N)}(t)| \sim  t^{N/2}$), before the center-of-mass magnetization $d(t)$ reaches zero, there is a nonuniversal region where the correlations start to build up and the amplitude $|c_\up^{(1, N)}(t)|$ reaches a maximum with constructive interference ($\theta_\sigma(t) \approx \pm \pi$).  The evolution for $\nu_0 t/N \gtrsim 0.02$ of $c_\up^{(1, N)}(t)$ is instead universal, the results for different values of $N$ collapse onto each other  on rescaling the time axis according to $t\rightarrow t/N$.
This scaling is the same as the one found for the average magnetization, suggesting that the oscillations of the correlation  $c_\sigma^{(1, N)}(t)$ in this time window are associated to 
the spin-mixing dynamics.

\begin{figure}
    \centering
   \includegraphics[scale=0.45]{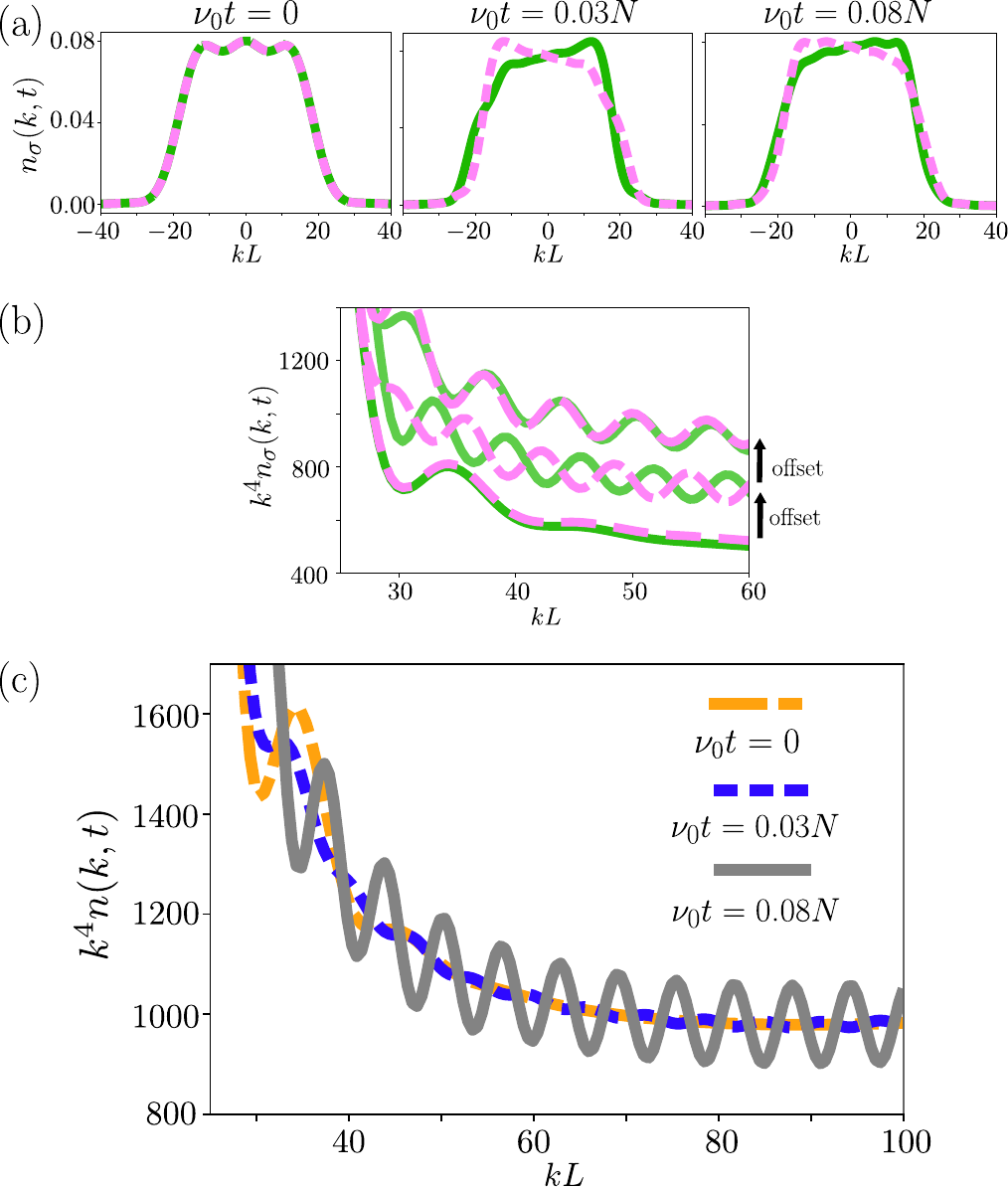}
    \caption{Dynamics of the single-component momentum distributions, $n_\sigma(k, t)$, for $N=6$ SU(2) fermions in a box trap at (a) low momenta and (b) large momenta  for $\sigma=\uparrow$ (solid green) and $\downarrow$ (dashed pink),  at three different times,  $\nu_0 t/N = 0, 0.03, 0.08$. To distinguish the curves in (b), we have added an offset of $200$ between the curves at different times, as indicated by the black arrow.  (c) Tail of the resulting total momentum distribution $n(k, t) = \sum_\sigma n_\sigma(k, t)$ at the same selected times, $\nu_0 t/N = 0, 0.03, 0.08$.}
    \label{fig:k4nk}
\end{figure}
   
The zero center-of-mass magnetization points are instead characterized by a region where the phase takes values around $\pi/2$, meaning that the two components oscillate in counter phase, as shown in Fig.~\ref{fig:dt_sig}(d) (see also Appendix~\ref{app:d0}). When $d(t)\approx 0$, the dominant contributions to $c_\up^{(1, N)}(t)$
come from terms involving the spin configuration of the initial state and its parity-reversed counterpart, namely
   $c_\up^{(1, N)}(t)|_{d(t)\approx 0} \approx \braket{\chi(t) | \up \up \dots \up \down \dots \down\down}\braket{ \up \dots \up \down \dots \down\down\up | \chi(t)} + \braket{\chi(t) | \up \down\down \dots \down \up \dots \up}\braket{ \down \down\dots \down \up \dots \up\up | \chi(t)}$.
This occurs because when $d(t)=0$ spin configurations that are parity transforms of each other must be equally populated. Finally, when the $\up$-component density move to the opposite side, we find again a minimum for $|c_\up^{(1, N)}(t)|$ and the phase is $\pi$. The same behavior is repeated until the center-of-mass magnetization $d(t)$ approaches a zero-value plateau, indicating the system has become fully mixed.

Additionally, we observe that the amplitude $|c_\up^{(1, N)}(t)|$ depends slightly on $N$, following a scaling law that varies with the initial state. For the single spin-domain initial state, it scales as $N^{-1/4}$. If we consider the amplitude of the oscillations [see Eq.~\eqref{eq:KN_t_sig}], then we have $\mathcal{A}_N N_\sigma |c_\up^{(1, N)}(t)| \sim (N^2/L^3) N^{-1/4} \sim N^{-5/4}$, which approaches zero in the thermodynamic limit, as expected.

To illustrate the behavior at the tails of the total momentum distribution, $n(k, t)=n_\uparrow(k, t)+n_\downarrow(k, t)$, we focus on the case with $N=6$ at three distinct times: $t=0$, $\nu_0 t/N = 0.03$ (when $d(t) \approx 0$) and $\nu_0 t/N = 0.08$ (when $d(t)$ is about to reach the first maximum).  Figures~\ref{fig:k4nk}(a)-(b) show how the momentum distribution for the single components is not constant in time  at small and large momenta, where momentum oscillations with amplitude $|c^{(1, N)}_\sigma(t)|$ develop at $t>0$. As shown in Fig.~\ref{fig:k4nk}(c), initially $k^4n(k, 0) = \mathcal{B}_N + \mathcal{C}_N$, since there is no correlation between the two sides of the trap. Interestingly, when $d(t) \approx 0$, the total correlation $c^{(1, N)}(t) =c^{(1, N)}_\uparrow(t)+c^{(1, N)}_\downarrow(t) $ vanishes and there are no momentum oscillations in $n(k, t)$, even though each component $c^{(1, N)}_\sigma(t)$ is nonvanishing, due to the counter-phase oscillations of the $\uparrow$ and $\downarrow$ components [see Figs.~\ref{fig:k4nk}(b)-(c)]. Finally, near the first peak of $d(t)$, both single-component and total spin correlations reach their maxima, coinciding with the point at which the total momentum distribution oscillations are maximal. 

The results of this section show that the study of the oscillation dynamics of the momentum distribution of a mixture trapped in a box potential provide a very detailed real-time information of the buildup, evolution, and fate of the border-to-border string correlation.

\section{Conclusion}
\label{sec:end}
In conclusion, we have studied the nonequilibrium dynamics of 1D strongly repulsive cold atomic gases confined in a hard-wall potential, focusing on balanced two-component fermionic spin mixtures.
The rigid boundaries of the box potential, achievable
in current experiments \cite{art:navon_box}, induce momentum oscillations in the tails of the momentum distribution. The amplitude of these oscillations reflects the dynamical buildup of border-to-border 
spin correlations, while their phase shift is dictated by the symmetries of the initial state.

We demonstrated that the initial growth of this high-order coherence term is highly sensitive to the nature of the initial state, particularly for extreme configurations such as spin domain walls and Néel states. The time evolution of these spin correlations provides a valuable tool for tracking the system dynamics, allowing us to connect the behavior of spin coherence with other observables, such as center-of-mass oscillations of the magnetization. The shared timescale governing both quantities highlights the intricate link between local and nonlocal correlations in these systems.

Our findings open up possibilities for controlling spin states with specific degrees of stringlike correlations~\cite{art:denNijs_SO, art:Kennedy_SO}. To obtain a precise targeted state, one could think to prepare the system in a certain configuration at finite but strong interactions, let it evolve,  and subsequently freeze its dynamics at a precise time by further increasing the repulsive interactions, for example via Feshbach resonances.  These time-dependent spin correlations could be of great relevance for future quench spectroscopy experiments \cite{art:quench_spec_Sanchez2019, art:bocini2024_roscilde}, offering a pathway to probe and engineer complex quantum states with tailored correlations.

\begin{acknowledgements}
 We acknowledge financial support from the ANR-21-CE47-0009 Quantum-SOPHA project.
\end{acknowledgements}

\appendix

\section{Proof of Eq.~\eqref{eq:KN_t_sig}}
\label{app:k4nk}

In this Appendix, we derive Eq.~\eqref{eq:KN_t_sig} by extending 
the derivation of  Ref.~\cite{art:aupetit2023} to wave functions with a time-dependent spin part [see Eq.~\eqref{eq:Psi_MB}]. 
We start by separating the contribution in $x$ and $x'$ of Eq.~\eqref{eq:nk_sigma}  as follows
\begin{equation}
\begin{split}
n_\sigma(k, t) &=  \sum_{i, j=1}^N (-1)^{i+j} c^{(i, j)}_\sigma (t) R^{(i, j)}(k) \\
&=\frac{N!}{2\pi} \sum_{i, j=1}^N (-1)^{i+j}  c^{(i, j)}_\sigma (t) \int_{I_{ij}} \prod_{n \neq i} 
dx_n
\\&\int_{x_{i-1}}^{x_{i+1}} dx e^{ik x} \Psi^\ast_A(\dots, x_{i-1}, x, x_{i+1}, \dots)\\
&\times \int_{x_{j-1}}^{x_j} dx' e^{-ik x'}  \Psi_A(\dots, x_{i-1}, x', x_{i+1}, \dots),
\end{split}
\label{eq:nksig_2int} 
\end{equation}
where $R^{(i, j)}(k)$ are the Fourier transforms of $\rho^{(i, j)}(x, x')$ [Eq.~\eqref{eq:rhoij}] in the relative coordinate $x-x'$. As explained in Sec.~\ref{sec:corr}, to study the large $k$ behavior of Eq.~\eqref{eq:nksig_2int}, we have to sum all the contributions given by the singular points of the one-body density matrix in real space.

\subsection{Edge terms}\label{subsec:borders}
We first consider the singularities around the edge of the trap, where the  wave function can be expanded as in Eq.~\eqref{eq:PsiA_exp_single}.  We then distinguish the two limiting cases, when we take the limits of $x$ and $x'$ to the same side, or to opposite sides. Because these limits affects the coordinates $x$ and $x'$, by definition also the indices $i$ and $j$ in Eq.~\eqref{eq:nksig_2int} have to be modified, respectively. The limits $x, x' \to L/2$ [corresponding to $(i, j) \to (N, N)$] and  $x, x' \to -L/2$ [corresponding to $(i, j) \to (1, 1)$] yield to two identical contributions, summarily
 \begin{equation}
 \begin{split}
 &c^{(1, 1)}_\sigma (t) R^{(1, 1)}(k) + c^{(N, N)}_\sigma (t) R^{(N, N)}(k) \\
 &\stackrel{k \to \infty}{=}  \frac{N!}{2 \pi} \frac{|\Phi_L(x_2, \dots, x_N)|^2}{k^4} \left(c^{(1, 1)}_\sigma (t) + c^{(N, N)}_\sigma (t) \right)\\
 &\equiv \frac{\mathcal{B}_N}{k^4}\left(c^{(1, 1)}_\sigma (t) + c^{(N, N)}_\sigma (t) \right),
 \end{split}
 \label{eq:Rij_B}
 \end{equation}
where we have introduced the coefficient $\mathcal{B}_N$ of Eq.~\eqref{eq:KN_t_sig}. For an initial state which conserves time-reversed symmetry, one can use $c_\sigma^{(i, i)}(t) = 1 - c_\sigma^{(N+1-i, N+1-i)}(t)$  and, therefore, obtain a time independent term.   
In addition, when we used the Watson lemma, we have divided by two because of the half cusp and the two exponentials from the Watson lemma have opposite sign $e^{\pm ikL/2}\times e^{\mp ikL/2} = 1$.
 
 Instead, the limits  $x\to L/2$ and $x'\to -L/2$ [corresponding to $(i, j) \to (N, 1)$] or   $x\to -L/2$ and $x'\to L/2$ [corresponding to $(i, j) \to (1, N)$] give the contribution
  \begin{equation}
 \begin{split}
 &(-1)^{N+1} \Big(c^{(1, N)}_\sigma (t) R^{(1, N)}(k) + c^{(N, 1)}_\sigma (t) R^{(N, 1)}(k) \Big)\\
 &\stackrel{k \to \infty}{=}  -\frac{ N!}{2\pi}\left(c^{(1, N)}_\sigma (t)e^{-ikL} + c^{(N, 1)}_\sigma (t)e^{ikL}  \right)\\
 &\times\frac{|\Phi_L(x_2, \dots, x_N)|^2}{k^4}
 \\
 &\equiv -\frac{\mathcal{A}_N}{k^4} |c^{(1, N)}_\sigma (t)|2 \cos(kL+\theta_\sigma(t)), 
 \end{split}
 \label{eq:Rij_A}
 \end{equation}
 where we have used that the remaining contribution given by the regular part of the wave function [$\Phi_L$ in Eq.~\eqref{eq:PsiA_exp_single}] is the same for the two cases (the coordinates in the middle are ordered in the same way: $x_2 < x_3 < \cdots < x_N$) and divided by two because of the half cusp. In the last line, 
$c^{(1, N)}_\sigma (t) = [c^{(N, 1)}_\sigma (t)]^\ast$ are complex coefficients and $\theta_\sigma(t)$ is a time-dependent phase. In the presence of time-reversal symmetry, we have also that  $c^{(1, N)}_\up (t) = [c^{(1, N)}_\down (t)]^\ast$ and this implies that $\theta_\up (t) = - \theta_\down(t)$.  Finally, because we consider only equally balanced mixture the factor $(-1)^{N+1} = -1$.
 
We now calculate the coefficients in front of the $1/k^4$ in Eq.~\eqref{eq:Rij_B} and~\eqref{eq:Rij_A}. Following Ref.~\cite{art:deuretzbacher2016_num}, we use the Leibniz formula for a determinant such that $\Psi_F =(1/\sqrt{N!})\sum_{P\in S_N} \epsilon(P) \prod_{i=1}^{N} \phi_{P(i)}(x_i)$,
where $\epsilon(P)$ is the signature of the permutation $P$, we rewrite the spatial part of  Eq.~\eqref{eq:rho1} in terms of the overlap integrals $A_{p,q}(z) \equiv \int_{z}^{L/2}du~\phi_{p}(u)\phi_{q}(u)$ as follows:
\begin{equation}
\begin{split}
    \rho^{(i,j)}(x,x')&=\dfrac{1}{(i-1)!(j-i)!(N-j)!}\sum_{P,Q\in S_N}\epsilon(P)\\
    &\times\epsilon(Q)\phi_{P(1)}(x)\phi_{Q(1)}(x')
    \\&\times\prod_{k=2}^{i}( \delta_{P(k), Q(k)}-A_{P(k),Q(k)}(x))
    \\&\times \prod_{l=i+1}^{j}(A_{P(l),Q(l)}(x)-A_{P(l),Q(l)}(x'))\\
    &\times \prod_{m=j+1}^{N}A_{P(m),Q(m)}(x'),
\end{split}
 \label{eq:rhoij_Aij}
\end{equation}
which is valid for $x \leq x'$. For the box potential,  the single-particle orbitals are $\phi_n(x)=\sqrt{2/L}\,\sin{[(n\pi/L) (x+L/2)]}$ with   $n\in\{1,\dots,N\}$ and the overlap integrals are given by
\begin{equation}
\begin{split}
A_{p,q}(z)&=\dfrac{\sin\left(\dfrac{\pi}{2L}(p+q)(2z+L)\right)}{\pi(p+q)}\\
&-\dfrac{\sin\left(\dfrac{\pi}{2L}(p-q)(2z+L)\right)}{\pi(p-q)}.
\end{split}
\label{eq:Apq}
\end{equation}

As mentioned before, for the limits $x, x' \to \pm L/2$, the only terms that remain are those with $(i, j) = (1, 1)$ and $(i, j) = (N, N)$, respectively, whose spatial parts are the same for a symmetric trap (parity is conserved). Therefore, from Eq.~\eqref{eq:rhoij_Aij}, we can consider only one term, such as
\begin{equation}
\begin{split}
    \rho^{(1,1)}(x,x')&=\dfrac{1}{(N-1)!}\sum_{P,Q\in S_N}\epsilon(P)\epsilon(Q)\phi_{P(1)}(x) \\
    &\times \phi_{Q(1)}(x') \prod_{k=2}^{N}A_{P(k),Q(k)}(x),
    \end{split}
    \label{eq:rhoNN}
\end{equation}
and $\forall p, q$
\begin{align}
    &\lim_{\substack{x\rightarrow \pm \frac{L}{2}\\ x'\rightarrow \pm\frac{L}{2}}}\phi_{p}(x)\phi_{q}(x')=\dfrac{2\pi^2}{L^3}pq\left(x \mp \dfrac{L}{2}\right)\left(x' \mp \dfrac{L}{2}\right),\label{Xlim cte}\\
    &\lim_{\substack{z\rightarrow \frac{L}{2}}} A_{p,q}(z)=0,\quad \text{and} \quad \lim_{\substack{z\rightarrow -\frac{L}{2}}} A_{p,q}(z)= \delta_{p, q},\label{Alim}
\end{align}
where we can see that the single-particle orbitals give the singular behavior of Eq.~\eqref{eq:PsiA_exp_single}. 
Therefore, by following Ref.~\cite{art:aupetit2023} and applying the Watson lemma for the half cusp, we define
\begin{equation}
\begin{split}
   \mathcal{B}_N &= \frac{1}{2\pi} \left(\frac{2\pi^2}{L^3} \sum_{P} P(1)^2\right) = \frac{\pi}{L^3} \frac{(N+1)(2N+1)}{6}.
    \end{split}
    \label{eq:BN}
\end{equation}

We now consider the limits $(x, x') \to (\pm L/2, \mp L/2)$, for which only the terms $(1, N)$ and $(N, 1)$ are nonzero. The spatial part is also the same in this case, namely $\rho^{(1, N)}(x, x') = \rho^{(N, 1)}(x', x)$. Therefore, from Eq.~\eqref{eq:rhoij_Aij}, we find that 
\begin{equation}
\begin{split}
    \rho^{(1,N)}(x,x')&=\dfrac{1}{(N-1)!}\sum_{P,Q\in S_N}\!\!\!\epsilon(P)\epsilon(Q)\phi_{P(1)}(x)\phi_{Q(1)}(x') \\&\times\prod_{l=2}^{N}(A_{P(l),Q(l)}(x)-A_{P(l),Q(l)}(x')),
 \end{split}
 \label{eq:rho1N}
\end{equation}
and we notice that
\begin{align}
    &\lim_{\substack{x\rightarrow -\frac{L}{2}\\ x'\rightarrow \frac{L}{2}}}\phi_{p}(x)\phi_{q}(x')=\dfrac{2\pi^2}{L^3}(-1)^qpq\left(x+\dfrac{1}{2}\right)\left(x'-\dfrac{1}{2}\right),\label{Xlim osc}\\
    &\lim_{\substack{x\rightarrow -\frac{L}{2}\\ x'\rightarrow \frac{L}{2}}}\left(A_{p,q}(x')-A_{p,q}(x')\right)=\delta_{p,q}.\label{AmAlim}
\end{align}
Therefore, as in Eq.~\eqref{eq:BN}, we define the coefficient in front of the $1/k^4$ of Eq.~\eqref{eq:Rij_A} as
\begin{equation}
\begin{split}
   \mathcal{A}_N &= \frac{1}{2\pi} \left(\frac{2\pi^2}{L^3} \sum_{P} (-1)^{P(1)} P(1)^2\right) \\
   &= \frac{\pi}{2 L^3} (N+1) (-1)^N.
    \end{split}
    \label{eq:AN}
\end{equation}

\subsection{Short-distance term}\label{subsec:shortdis}
We now focus on the short-distance limit $x \to x'$, which corresponds
to $(i, j)  \to (i, i)$ or $(i, i+1)$ and gives the following contribution:
  \begin{equation}
 \begin{split}
 &\sum_{i=1}^N \sum_{j = i, i+1} (-1)^{i+j} c^{(i, j)}_\sigma (t) R^{(i, j)}(k) \\
 &\stackrel{k \to \infty}{=}\frac{2 N!}{\pi} \sum_{i=1}^{N-1} \left(c^{(i, i)}_\sigma(t)- c^{(i, i+1)}_\sigma(t)\right) \\
&\times\sum_{l \neq i} \int_{K_{i}}\left(\prod_{n\neq i}^N dx_n \right)
 \frac{|\Phi_0 (x, x_{n\neq i, l})|^2}{k^4},
 \\
 &\equiv \frac{\mathcal{C}_N^\sigma (t)}{k^4},
 \end{split}
 \label{eq:Rij_Tan}
 \end{equation}
where we have used Eq.~\eqref{eq:PsiA_exp_single} of the main text and  $K_i = \{x_1 < \dots < x_{i-1} < x \leq x' < x_{i+1} < \dots < x_N\}$ such that the only $j$ that survives in the sum are $j=i$ ($x=x'$) and $j=i+1$ ($x \lesssim  x'$). To obtain the second line of Eq.~\eqref{eq:Rij_Tan}, we have used that~\cite{art:musolino2024, art:Patu2017}
\begin{equation}
\begin{split}
& \int_{I_{ij}}\left(\prod_{n\neq i}^N dx_n \right) \sum_{l, m\neq i}
\Phi_0^\ast (x, x_{n\neq i, l}) \Phi_0 (x', x_{n\neq i, m})  \\
&\times e^{-i k (x_m-x_l)} \\
&\approx \sum_{l \neq i} \int_{K_{i}}\left(\prod_{n\neq i}^N dx_n \right)
 |\Phi_0 (x, x_{n\neq i, l})|^2 = \frac{m^2 \alpha_i}{2N! \hbar^4},
\end{split}
\label{eq:nksigma_fin}
\end{equation}
where the off diagonal terms ($l\neq m$), which are multiplied by a product of regular functions, $\Phi_0^\ast \Phi_0$, are neglected because they vanish faster than $1/k^4$ due to Riemann-Lebesgue lemma~\cite{art:Patu2017}. In the last equality of Eq.~\eqref{eq:nksigma_fin}, we have introduced the coefficients $\alpha_i$, as defined in Eq.~\eqref{eq:Ji}, and used the cusp condition~\cite{art:deuretzbacher2014}, namely 
\begin{equation}
\begin{split}
\frac{mg}{\hbar^2} \Psi_A|_{x_i=x_l} &= \frac{\partial \Psi_A}{\partial (x_{i}-x_{l})}\Big|^{0+}_{0-}
\\&\stackrel{|x_i-x_l|\to 0}{\approx}  2 \Phi_0\left(x_i, x_{n\neq i, l}\right).
\end{split}
\label{eq:cusp_rel}
\end{equation}

We now demonstrate that the sum over $\sigma$ of $\mathcal{C}_N^\sigma(t)$ gives the Tan contact, $\mathcal{C}_N$,  which is defined in Eq.~\eqref{eq:Tan_def}.  To do so, we consider the spin Hamiltonian in Eq.~\eqref{eq:Hspin}, where the spatial part of the wave function is already integrated out. Therefore, one can write 
\begin{equation}
\begin{split}
\mathcal{C}_N(t) &= - \frac{m^2}{\pi \hbar^4}\braket{ \chi(t) |\frac{\partial \hat{H}_s}{\partial 1/g} | \chi(t)} \\&=
  -\frac{m^2}{\pi \hbar^4} \sum_{i=1}^{N-1} 2\alpha_i  \bra{\chi(t)} \Big(\mathbf{S}_{i}\cdot \mathbf{S}_{i+1}  -\frac{1}{4} \Big)\ket{\chi(t)}
  \\
  &=\frac{m^2}{\pi \hbar^4} \sum_{i=1}^{N-1} \alpha_i  \braket{\chi(t)| \left( 1 - \hat{P}_{i, i+1}\right)| \chi(t)}\\
   &= \frac{m^2}{\pi \hbar^4} \sum_{i=1}^{N-1} \alpha_i \left( c^{(i, i)}(t) - c^{(i,i+1)}(t)\right),
   \end{split}
   \label{eq:adiab_time}
\end{equation}
where, in the second line, we have used the relation $\hat{P}_{i, i+1} = 2 \mathbf{S}_i \cdot \mathbf{S}_{i+1} + 1/2$~\cite{art:zhang_wang}.  By separating the two $\sigma$ components and using Eq.~\eqref{eq:nksigma_fin}, we find $\mathcal{C}_N = \sum_\sigma \mathcal{C}_N^\sigma (t)$. We notice that we have a minus sign in front of the $(i, i+1)$ correlation instead of a plus sign for bosonic mixtures (see the supplemental material of Ref.~\cite{art:musolino2024}).

Finally, we show that $\mathcal{C}_N$ does not depend on time for the nonequilibrium scenario studied in this paper. Although the $i$th coefficients, $c^{(i, i)}(t)$ and $c^{(i, i+1)}(t)$, are time dependent, the sum over $i$ is not. This can be shown by considering the third line of Eq.~\eqref{eq:adiab_time}
 \begin{equation}
 \begin{split}
 \mathcal{C}_N &= \frac{m^2}{\pi \hbar^4}\sum_{i = 1}^{N-1} \alpha_i \braket{\chi(t) | (1 - \hat{P}_{i, i+1}) |\chi(t)} \\
 &= \frac{m^2 g}{\pi \hbar^4} \braket{\chi(t) | \hat{H}_s |\chi(t)}\\
 &= \frac{m^2 g}{\pi \hbar^4}\sum_{m, n} s_m s_n \braket{\chi_m | \hat{H}_s|\chi_n} e^{-i(E_n - E_m)t}\\
 &=  \frac{m^2 g}{\pi \hbar^4}\sum_{n} s_n^2 E_n, 
 \end{split}
 \label{eq:C_spin}
 \end{equation}
where we have neglected the constant term of the Hamiltonian [Eq.~\eqref{eq:Hspin}], which gives only an offset, and decomposed the spin states in the basis that diagonalizes the Hamiltonian, $\ket{\chi(t)} = \sum_n s_n  e^{-iE_n t}\ket{\chi_n}$.

\section{Short-time dynamics of the spin correlations}
\label{app:corr_tshort}
\begin{figure}[h]
   \includegraphics[scale=1]{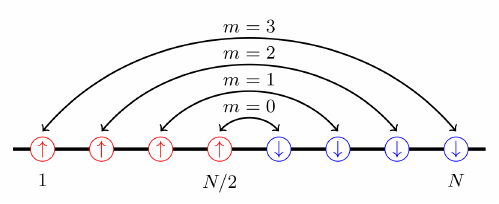}
\caption{Illustration of the spin correlations  $c^{(N/2-m, N/2+1+m)}_\sigma(t)$ for $N=8$ for a state with two fully separated components (red, $\up$ and blue, $\down$).}
\label{fig:corr}
\end{figure}

In this section, we prove that for the initial state $\ket{\chi(0)}= \ket{\up \up \dots \up \down \dots \down \down}$, the spin correlations of the type  $c^{(N/2-m, N/2+1+m)}_\sigma (t)$ at short times behaves as  $t^{m+1}$  with $0 \leq m \leq N/2-1$, where the limiting cases are the correlation at the interface ($m=0$), $c^{(N/2, N/2+1)}_\sigma (t)$, and the correlation border to border ($m=N/2-1$), $c^{(1, N)}_\sigma(t)$. The latter has been discussed in Sec.~\ref{sec:HOSC}. As illustrated in Fig.~\ref{fig:corr}  for $N=8$, the different $c^{(N/2-m, N/2+1+m)}_\sigma (t)$  all correspond to correlations centered in the middle of the trap.

To study the short-time behavior of $c^{(N/2-m, N/2+1+m)}_\sigma (t)$ for  $\sigma=\uparrow$ and a generic $N$ and $m$, we start by writing the Taylor expansion of the two exponentials of Eq.~\eqref{eq:cij_single} around $t=0$, as follows
\begin{equation}
    \begin{split}
    &c_{\up}^{(\Ntwo-m, \Ntwo+1+m)}(t) = \sum_{\vec{\sigma} \in \Sigma} \delta_{\up, \sigma_{\Ntwo-m}} 
    \braket{\chi(0)|e^{i\hHs t} |\vec{\sigma}} \\
    &\bra{\vec{\sigma}}\hat{P}_{\left(\Ntwo-m, \dots, \Ntwo+1+m\right)}e^{-i\hHs t}\ket{\chi(0)} \\
    &\stackrel{t\approx 0}{=}\sum_{\vec{\sigma} \in \Sigma}\delta_{\up, \sigma_{\Ntwo-m}} 
    \braket{\chi(0)| \sum_{\beta_1=0}^\infty  \frac{(i\hHs t)^{\beta_1}}{\beta_1!}|\vec{\sigma}}\\ &\bra{\vec{\sigma}} \hat{P}_{\left(\Ntwo-m, \dots, \Ntwo+1+m\right)} \sum_{\beta_2=0}^\infty \frac{(-i\hHs t)^{\beta_2}}{\beta_2!}\ket{\chi(0)},
    \end{split}
    \label{eq:cm_sigma_tapprox0}
\end{equation} 
where the cycle $\hat{P}$ is made of $(2m+1)$ adjacent transpositions, as 
\begin{equation}
\begin{split}
\hat{P}_{\left(\Ntwo-m, \dots, \Ntwo+1+m\right)} &= \hat{P}_{\Ntwo-m, \Ntwo-m+1} \hat{P}_{\Ntwo-m+1, \Ntwo-m+2} 
\\&\cdots  \hat{P}_{\Ntwo+m, \Ntwo+m+1},
\end{split}
\label{eq:Pn2m}
\end{equation} 
between the spin at position $\Ntwo-m$ and the one at position $\Ntwo+1+m$ of the configuration. Because of the $\delta_{\up, \sigma_{\Ntwo-m}}$, the zero-order term ($\beta=\beta_1+\beta_2 = 0$) of Eq.~\eqref{eq:cm_sigma_tapprox0}  is zero. 

\begin{figure}
    \centering
   \includegraphics[scale=0.6]{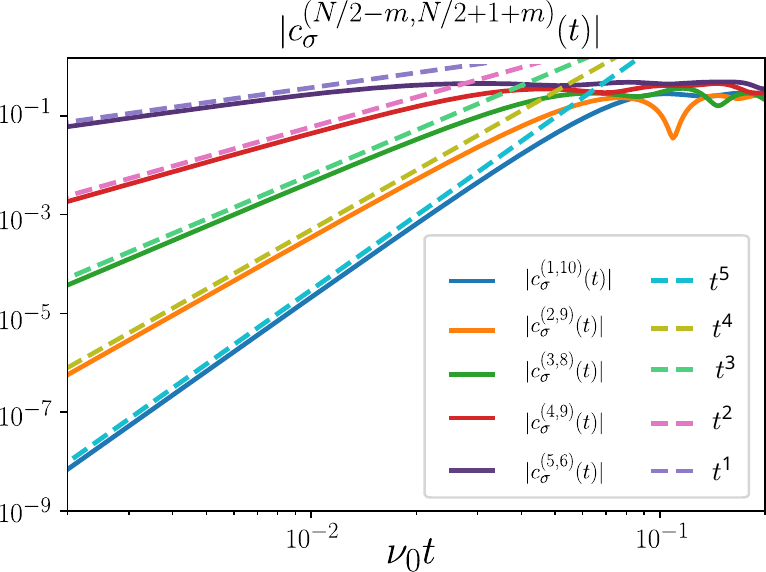}
    \caption{Initial growth of the magnitude of the single-component spin correlation $c^{(N/2-m, N/2+1+m)}_\sigma(t)$ for $N=10$ in log-log scale and the initial state $\ket{\up \up \cdots \up \down \cdots \down \down}$. The dashed lines represent the fitting functions that behave as $\sim t^{n_s}$ with $n_s$ the minimum number of exchanges necessary to bring the spin at position $N/2$ to position $N/2+1+m$.}
    \label{fig:cij_sigma_order}
\end{figure}

 We now show how many times at least we need to apply $\hHs$ to obtain a nonzero contribution on the right-hand side of Eq.~\eqref{eq:cm_sigma_tapprox0}, which can be written as
\begin{equation}
\begin{split}
&\delta_{\up, \sigma_{\Ntwo-m}} \bra{\vec{\sigma}} \hat{P}_{\left(\Ntwo-m, \dots, \Ntwo+1+m\right)}  \Big(\sum_{i=1}^{N-1} \hat{P}_{i, i+1} \Big)^{\beta_2-1} \\
&\hat{P}_{\Ntwo, \Ntwo+1}\ket{\up \cdots \up \down \cdots \down }\\
&= \delta_{\up, \sigma_{\Ntwo-m}} \bra{\vec{\sigma}}\hat{P}_{\left(\Ntwo-m, \dots, \Ntwo+1+m\right)}  \Big(\sum_{i=1}^{N-1} \hat{P}_{i, i+1} \Big)^{\beta_2-2} \\
&\Big( \hP_{\Ntwo-1, \Ntwo} + \hP_{\Ntwo, \Ntwo+1} + \hP_{\Ntwo+1, \Ntwo+2} \Big) \ket{\up \cdots \up \down \up \down \cdots \down},
\end{split}
    \label{eq:alphaorder_m}
\end{equation}
where the first time that we apply $\hHs$ on the initial state, only the transposition of the spin at the interface, $P_{\Ntwo, \Ntwo+1}$, modifies the initial state,  while the second time, we can only take into account the transpositions that involve the spins surrounding the interface, and so on.  By inspection, to obtain a nonzero result, we must bring the spin $\up$ at position $N/2-m$ to the position $N/2+1+m$ such that the $(N/2-m, \dots, N/2+1+m)$ cycle will bring it back to the initial position. 
This means that we need to apply the Hamiltonian at least $\beta_2=m+1$ times.  Finally, the lowest order of the left-hand Taylor expansion  of Eq.~\eqref{eq:cm_sigma_tapprox0} that contributes  is the zero order, $\beta_1=0$. Therefore, we conclude that the power law that matters at the shortest times is $t^{m+1}$ with $\beta_1=0$ and $\beta_2=m+1$. We notice that this result can be generalized to other trapping potentials by introducing the inhomogeneous coefficients $J_i$ in $\hat H_s$~\cite{art:deuretzbacher2014}. Figure~\ref{fig:cij_sigma_order} shows the initial growth of the magnitude of $c^{(N/2-m, N/2+1+m)}_\sigma (t)$ with $m=0, 1, 2, 3, 4$ for $N=10$ and, as expected, we find the  $t^{m+1}$ growth at small times.

\subsection{Examples}
To give some explicit examples of the proof presented above, here we focus on the short-time behavior of $c_\sigma^{(1, N)}(t)$ ($m=N/2-1$) for  $N=4$ and $N=6$, which are shown in Fig.~\ref{fig:c1N_sigma}.  We start with $N=4$. While it is straightforward to verify that the zero and first order of Eq.~\eqref{eq:alphaorder_m} are zero, the second order  reads as 
\begin{equation}
\begin{split}
& \delta_{\up, \sigma_{1}}  \bra{\vec{\sigma}}\hP_{1,2} \hP_{2,3}\hP_{3,4}  \Big( \hP_{1, 2}+ \hP_{2, 3} + \hP_{3, 4} \Big) \hP_{2, 3} \ket{\up \up \down  \down },
\end{split}
    \label{eq:2order_N4}
\end{equation}
where the only term that survives is the third one (using that  $\hP_{1,2} \hP_{2,3}\hP_{3,4} \hP_{3, 4} \hP_{2, 3} =\hP_{1,2} $ and $\hP_{i, j} \hP_{i, j} = \mathrm{Id}$) and therefore $c_{\sigma}^{(1, 4)}(t\approx 0) \sim t^2$.

Analogously, for $N=6$, the zero and first order for $c_\sigma^{(1, N)}(t)$ are straightforwardly zero, but also the second order goes to zero. Indeed, from Eq.~\eqref{eq:alphaorder_m}, we can write 
\begin{equation}
\begin{split}
& \delta_{\up, \sigma_{1}}\bra{\vec{\sigma}} \hP_{1,2} \hP_{2,3}\hP_{3,4}\hP_{4,5}\hP_{5,6}  \Big( \hP_{2, 3}+ \hP_{3, 4} + \hP_{4, 5} \Big) \\
&\hP_{3, 4}\ket{\up \up \up \down  \down \down},
\end{split}
    \label{eq:2order_N6}
\end{equation}
which is always zero, because a spin $\up$ never reaches the position $N$. The third order instead gives 
\begin{equation}
\begin{split}
& \delta_{\up, \sigma_{1}}\bra{\vec{\sigma}} \hP_{1,2} \hP_{2,3}\hP_{3,4}\hP_{4,5}\hP_{5,6} \Big( \hP_{1, 2}+ \hP_{2, 3} + \hP_{3, 4} \\
& + \hP_{4, 5}+ \hP_{5, 6}\Big)  \Big( \hP_{2, 3}+ \hP_{3, 4} + \hP_{4, 5} \Big) \hP_{3, 4}\ket{\up \up \up \down  \down \down},
\end{split}
    \label{eq:3order_N6}
\end{equation}
where the term $\hP_{1,2} \hP_{2,3}\hP_{3,4}\hP_{4,5}\hP_{5,6}\hP_{5,6}  \hP_{4, 5} \hP_{3, 4} = \hP_{1,2} \hP_{2,3}$ gives back the initial state and therefore a nonzero contribution at the third order, namely $c_{\sigma}^{(1, 6)}(t\approx 0) \sim t^3$, as expected.

\subsection{Initial state with N\'eel order}
\begin{figure}[h]
    \centering
   \includegraphics[scale=0.7]{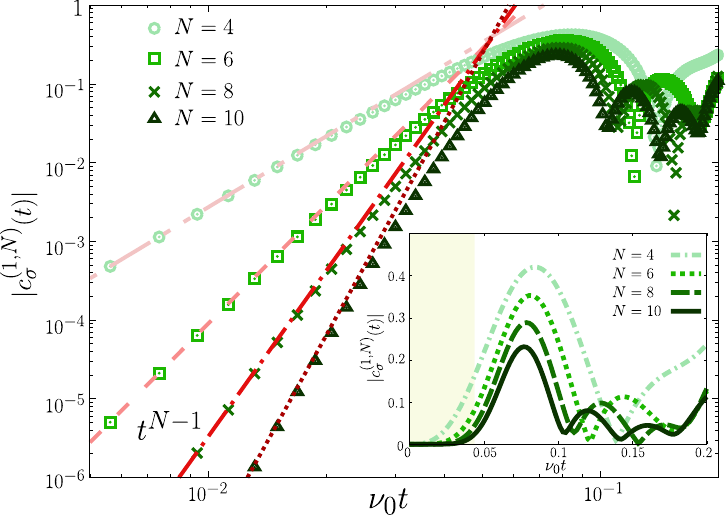}
\caption{Initial growth of the modulus of the $c_\sigma^{(1, N)}(t)$ correlations for the initial state $\ket{\chi(0)} =\ket{\uparrow\downarrow\uparrow\downarrow \dots \uparrow\downarrow}$ for $4 \leq N \leq 10$ in log-log scale. The fitting functions correspond to $t^{N-1}$. Inset: Same as the main figure without the log-log scale. The yellow area indicates the region where the power-law $t^{N-1}$ is a good fit. }
\label{fig:c1N_Neel}
\end{figure}

From the proofs above, it is clear that the spin composition of the initial state plays an important role on the short-time dynamics of the spin correlations. In this section, we discuss the initial state with alternated spins or with N\'eel order, namely $\ket{\up \down \up \down \dots \up \down \up \down}$. As mentioned in Sec.~\ref{sec:HOSC}, for this state, we have found that $c_\sigma^{(1, N)} (t) \sim t^{N-1}$ at small times. This result has been numerically checked for $N \leq 10$, as shown in Fig.~\ref{fig:c1N_Neel}.    

To demonstrate that, for simplicity, we only consider $N=4$. It is straightforward to verify that the zero and first order are zero because $\delta_{\up, \sigma_1}\bra{\vec{\sigma}}P_{(1, \dots, 4)} \ket{\up \down \up \down} = \delta_{\up, \sigma_1}\braket{\vec{\sigma}|\down \up \down \up} = 0$ and
\begin{equation}
\begin{split}
&\delta_{\up, \sigma_1}\bra{\vec{\sigma}}\hat{P}_{(1, \dots, 4)} \hat{H}_s \ket{\up \down \up \down} = \delta_{\up, \sigma_1}\bra{\vec{\sigma}} \Big(\ket{\down \down \up \up} \\
&+ \ket{\down  \up \up \down} + \ket{ \up \up \down \down}\Big)
\end{split}
\label{eq:P14H}
\end{equation}
where the only nonzero contribution gives $\ket{\vec{\sigma}} = \ket{ \up \up \down \down}$ which cannot be obtained on the left Taylor expansion for $\beta_1=0$.  
However, the latter contributes at the second order by combining the first orders of the two Taylor expansions, $\beta_1=1$ and $\beta_2=1$, namely  $ \braket{\up \down \up \down|\hat{P}_{2,3}| \vec{\sigma}}\braket{\vec{\sigma} |\hat{P}_{(1, \dots, 4)} \hat{H}_s| \up \down \up \down}t^2 = t^2$, where $\hat{P}_{2,3}$ comes from the Hamiltonian. 
In addition, at the second order, we have the following term for $\beta_1=0$ and $\beta_2=2$:
\begin{equation}
\begin{split}
&-\braket{\up \down \up \down| \vec{\sigma}}\braket{\vec{\sigma} |\hat{P}_{(1, \dots, 4)} \hat{H}_s^2|\up \down \up \down}\frac{t^2}{2}\\
&=-\braket{\up \down \up \down| \vec{\sigma}}\braket{\vec{\sigma} |\hat{P}_{(1, \dots, 4)} (\hat{P}_{1,2}\hat{P}_{3,4}+ \hat{P}_{3,4}\hat{P}_{1,2})|\up \down \up \down}\frac{t^2}{2}\\
&=-t^2,
\end{split}
    \label{eq:P14H2}
\end{equation}
where in the second step we have kept only the terms that give $\hat{P}_{(1, \dots, 4)} \hat{H}^2_s\ket{\up \down \up \down} =  \hat{P}_{(1, \dots, 4)}\ket{ \down \up  \down \up} =\ket{\up \down \up \down}$. The two second-order terms cancel each other and  therefore, also the second order is zero. 

At the third order, we have also two terms. The first one with $\beta_1=1$ and $\beta_2=2$ reads as 
\begin{equation}
\begin{split}
&-i  \braket{\up \down \up \down| \hat{H}_s|\vec{\sigma}} \braket{\vec{\sigma} |\hat{P}_{(1, \dots, 4)} \hat{H}^2_s|\up \down \up \down} \frac{t^3}{2}\\
&= -i \braket{\up \up \down \down|\vec{\sigma}} \braket{\vec{\sigma}|\hat{P}_{(1, \dots, 4)} \hat{P}_{2,3}\hat{P}_{3,4} |\up \down \up \down} \frac{t^3}{2},
\end{split}
    \label{eq:P14H2_3ord}
\end{equation}
where in the left expansion only $\hat{P}_{3,4}$ gives a nonzero result and therefore $\ket{\vec{\sigma}} = \ket{\up \up \down \down}$ and in the right expansion we have only kept the term that gives $\hat{P}_{(1, \dots, 4)} \hat{H}^2_s\ket{\up \down \up \down} =  \hat{P}_{(1, \dots, 4)}\ket{ \up \down  \down \up} =\ket{\up \up \down \down}$. The second term at the third order  with $\beta_1=0$ and $\beta_2=3$ is the following:
\begin{equation}
\begin{split}
&i  \braket{\up \down \up \down|\vec{\sigma}} \braket{\vec{\sigma} |\hat{P}_{(1, \dots, 4)} \hat{H}^3_s|\up \down \up \down} \frac{t^3}{6}
\\
&= i  \braket{\up \down \up \down|\vec{\sigma}} \bra{\vec{\sigma}} \hat{P}_{(1, \dots, 4)} (\hat{P}_{3,4}\hat{P}_{2,3} \hat{P}_{1,2}\\
&+ \hat{P}_{2,3}  \hat{P}_{1,2}\hat{P}_{3,4}) \ket{\up \down \up \down} \frac{t^3}{6} = i \frac{t^3}{3},
\end{split}
    \label{eq:P14H3}
\end{equation}
where in the second step we see that we can go back to the initial state not only by applying the identity $\hat{P}_{(1, \dots, 4)} \hat{P}_{3,4}\hat{P}_{2,3} \hat{P}_{1,2} = \hat{P}_{(1, \dots, 4)} \hat{P}_{(1, \dots, 4)}^{-1}=  \mathrm{Id}$, but also $\hat{P}_{(1, \dots, 4)}\hat{P}_{2,3} \hat{P}_{1,2} \hat{P}_{3,4}$. By summing the contributions in Eqs.~\eqref{eq:P14H2_3ord} and~\eqref{eq:P14H3}, we conclude that $c_\up^{(1, 4)} (t) \sim -i t^3/6$ at small times. 

Contrary to the spin separated state, for the N\'eel state, there are more terms of the two Taylor expansions of Eq.~\eqref{eq:cm_sigma_tapprox0} that are nonzero at lower order: however, those terms have opposite signs and therefore they cancel each other out. This leads to a higher-order power-law behavior,  which is a consequence of the higher number of spin domains in the initial state. 

\section{The $c^{(1,N)}_\sigma(t)$ coherence at $d(t)\approx 0$}
\label{app:d0}
In this section, we discuss additional details of the results presented in Sec.~\ref{sec:tlonger} for longer timescales.
For a two-component balanced mixture, the center of mass of the magnetization is zero [$d(t)\approx 0$] when the magnetization distribution is symmetric under a parity transformation. This means that  the spin configurations that are connected by a parity transformation ($L\leftrightarrow R$) are equally populated,
namely (i) $|a_P|^2=|a_{P(L\leftrightarrow R)}|^2$.
This implies that one can write $a_P=|a_P|e^{i\theta_P}$
and (ii) $a_{P(L\leftrightarrow R)}=|a_P|e^{-i\theta_P}$.
Moreover, the  initial state with a single-spin domain, used for the results presented in Sec.~\ref{sec:res}, is orthogonal to its parity-reversed: Because parity symmetry is conserved during the dynamics, this property holds at any time.
This means that 
\begin{equation}
    \sum_P a_P^*(t)a_{P(L\leftrightarrow R)}(t)=0,
\end{equation}
and, consequently, at $d(t)\approx 0$
\begin{equation}
\sum_{P} |a_P|^2\cos(2\theta_P)=0,
\end{equation}
that is verified for any population if (iii) $\theta_P=\pm\pi/4$. 

From the condition (i) we deduce that, at $d(t)\approx 0$, the  spin states
$\ket{\uparrow\uparrow\dots\uparrow\downarrow\dots\downarrow\downarrow}$ and $\ket{\downarrow\downarrow\dots\downarrow\uparrow\dots\uparrow\uparrow}$ have to be equally (or almost equally) populated, and analogously for the states
$\ket{\uparrow\dots\uparrow\downarrow\dots\downarrow\downarrow\uparrow}$ and $\ket{\uparrow\downarrow\downarrow\dots\uparrow\uparrow}$.

Thus, if the term $\braket{\chi(t) | \up \up \dots \up \down \dots \down\down}\allowbreak\braket{ \up \dots \up \down \dots \down\down\up | \chi(t)}$
is dominant in the coherence $c_\up^{(1, N)}(t)|_{d\approx 0}$, as expected at least at the first zero at $d(t)$ because of the choice of the initial state, then
the term $\braket{\chi(t) | \up \down\down \dots \down \up \dots \up}\braket{ \down \down\dots \down \up \dots \up\up | \chi(t)}$ has to be present too, with the same weight (from (i)) and the same phase (from (ii)).

From conditions (iii) we can deduce that such a phase has to be $\pm(\frac{\pi}{4}\pm \frac{\pi}{4})$, so that $c_\up^{(1, N)}(t)|_{d\approx 0}$ has to be real (the phase is zero) or imaginary (the phase is $\pm\pi/2$).
Indeed, we observe that in correspondence of the first three zeros of $d(t)$ the
single-component momentum-distribution tail oscillations have a dephasing of $\pi$,
while for the fourth zero, the dephasing is vanishing.

\bibliographystyle{apsrev4-2}
\bibliography{biblio}
\end{document}